\documentclass[10pt, journal, anonymous=true]{IEEEtran}
\usepackage{times}
\usepackage{latexsym}

\usepackage{microtype}
\usepackage{authblk}

\usepackage{amsmath}
\usepackage{amssymb}
\usepackage{url}
\usepackage{algorithmic}
\usepackage{comment}
\usepackage{subfig}

\usepackage{color}
\usepackage{xcolor}
\usepackage{tipa}
\usepackage{tikz}
\DeclareGraphicsExtensions{.pdf,.jpeg,.png}

\let\labelindent\relax
\usepackage{enumitem}
\usepackage{multirow}
\usepackage{pifont}% http://ctan.org/pkg/pifont
\usepackage{colortbl}
\usepackage{hhline}
\setitemize{noitemsep,topsep=0pt,parsep=0pt,partopsep=0pt}

\newlist{rqs}{enumerate}{1}
\setlist[rqs]{label*=\textbf{RQ\arabic*}}

\begin{document}

\title{Pitch Imperfect: Detecting Audio Deepfakes Through Acoustic Prosodic Analysis}
\author{Kevin Warren, Daniel Olszewski, Seth Layton, Kevin Butler,~\IEEEmembership{Senior Member,~IEEE}, Carrie Gates, and Patrick Traynor,~\IEEEmembership{Senior Member,~IEEE}}

\maketitle

\begin{abstract}
Audio deepfakes are increasingly indifferentiable from organic speech, often fooling both authentication systems and human listeners. While many techniques use low-level audio features or optimization black-box model training, focusing on the features that humans use to recognize speech will likely be a more long-term robust approach to detection. We explore the use of prosody, or the high-level linguistic features of human speech (e.g., pitch, intonation, jitter) as a more foundational means of detecting audio deepfakes. We develop a detector based on six classical prosodic features and demonstrate that our model performs as well as other baseline models used by the community to detect audio deepfakes with an accuracy of 93\% and an EER of 24.7\%. More importantly, we demonstrate the benefits of using a linguistic features-based approach over existing models by applying an adaptive adversary using an $L_{\infty}$ norm attack against the detectors and using attention mechansisms in our training for explainability. We show that we can explain the prosodic features that have highest impact on the model's decision (Jitter, Shimmer and Mean Fundamental Frequency) and that other models are extremely susceptible to simple $L_{\infty}$ norm attacks (99.3\% relative degradation in accuracy). While overall performance may be similar, we illustrate the robustness and explainability benefits to a prosody feature approach to audio deepfake detection. 
\end{abstract}


\section{Introduction} 
\label{sec:intro} 

Recent advances in audio deepfake generation techniques make creating human-sounding audio of anyone's voice more accessible and rapidly producible. While deepfakes can have enormous positive benefits~\cite{MBP14,leviathan_2018_google}, they also present the potential for misuse in cases such as fraud~\cite{FC23,JH23} and disinformation~\cite{SM23,VO23}. Society's exposure to deepfakes has grown in recent years as the success of deepfakes in social media has increased their visibility. In response, there has been a large effort by researchers to explore potential options for detecting such threats. Early deepfake detection techniques exploits black-box
machine learning (ML) techniques and low-level flaws (e.g., unusual spectral correlations~\cite{ALF19}, abnormal noise level estimations~\cite{pan2012detecting}, and unique cepstral patterns~\cite{balamurali2019toward}) to develop defenses. 

Recent work by Warren et al. studying the ways that human's percieve audio deepfakes demonstrates that humans and current machine learning models detect deepfakes differently and have separate trade-offs in performance~\cite{WTC_24}. They assert that these differences require that both humans and models are needed together for the detection process. Because of this, we take a different approach than the previous detection techniques and create a model that mimics the human's side in the detection process. In this paper, we explore a detection approach that focuses on the higher-level linguistc features that humans use when classifying deepfakes. We characterize {\it prosody}, the features of speech that are related to intonation, rhythm, and stress. While work has been done to improve prosody in deepfakes~\cite{sun_generating_2020, sun_fully-hierarchical_2020, skerry-ryan_towards_nodate, karlapati_prosodic_2021, hodari_camp_2021,chen_speech_2021, zhang_learning_2020,yang_exploiting_2020, fu_bi-level_2021, hono_hierarchical_2020, lancucki_fastpitch_2021, aggarwal_using_2020, valle_flowtron_2020}, characterizing prosody based on any single metric remains an open challenge in applied linguistics~\cite{SJP09} and faults in prosody are the most common reason people believe a piece of audio is a deepfake~\cite{WTC_24}. 

We explore the viability of prosodic features as a deepfake detection approach and highlight the benefits this approach has over existing models. Our efforts produce the following contributions:

\begin{itemize}

    \item {\bf Evaluate Prosody-Based Deepfake Detection:} We develop
	a classifier using six prosodic features (average and standard
	deviation of the mean-$F_0$, jitter, shimmer, average
	and standard deviation of harmonic to noise ratio (HNR)). We
	train/test our classifier on the ASVspoof2021
	dataset~\cite{yamagishi_asvspoof_2021} and show that we 
	discriminate deepfakes with 93\% accuracy. Additionally, we demonstrate 
    similar performance to baseline models across all standard ML performance
    metrics. 

    \item {\bf Provide Decision Explainability:} We implement an 
    attention mechanism into our training process to determine which prosodic features 
    have the largest influence on the model's classification. We show that three features 
    (e.g., jitter, shimmer, and mean-$F_0$) have the largest impact on the model's decision. 

    \item {\bf Characterize an Adaptive Adversary:}
	No prior work considers
	an adaptive adversary and its impacts on detection accuracy. We
	run an $L_{\infty}$ norm
	adversary~\cite{kurakin2016adversarial} against the best
	available baseline and show a 99.3\% \emph{decrease} in
	accuracy with minimal perturbation. 
	This demonstrates that miniscule amounts of white noise
	can bypass other detectors, while our approach is not susceptible 
	to such attacks.

\end{itemize}

This paper uses the linguistic characteristics of generated speech to aid in the detection of deepfakes. We anticipate that as audio deepfake generators improve and eliminate the flaws
that many current detectors build on, overcoming the task of mimicking prosody will be a far greater challenge. This work serves as the beginning of a long-term effort to consider the
intersection of linguistics and speech to protect against deepfake threats.

This paper is organized as follows:
Section~\ref{sec:relwork} presents
related work; Section~\ref{sec:acoustic} gives a background on prosody;
Section~\ref{sec:taxonomy} provides a deepfake taxomony; 
Section~\ref{sec:rqs} discusses our research questions; 
Section~\ref{sec:dataset} details the dataset used;
Section~\ref{sec:model} presents our model compared to 4 baseline detectors; 
Section~\ref{sec:adversarial} contextualizes our model decisions and tests against an adaptive adverary;
Section~\ref{sec:disc} discusses
related challenges; and Section~\ref{sec:conc} concludes.

\section{Related Work} 
\label{sec:relwork} 
 
The emergence and advancement of raw audio generation techniques have vastly improved the quality of audio attempting to sound organic/natural to the human ear~\cite{oord_wavenet_2016, oord_parallel_2017,baird2018the,kim_flowavenet_2019}.  Deepfake audio aims to impersonate real people to make the differentiation between deepfake and human speech
difficult~\cite{lorenzo-trueba_can_2018, SaundersDetectingDF}.  The potential for dangerous applications of fake audio has created the need for automated demarcation of humans from deepfakes. 

Audio deepfake detection was associated with spoof detection for automatic speaker verification (ASV) systems and spawned challenges such as ASVspoof2015~\cite{wu_asvspoof_2015} and ASVspoof2019~\cite{todisco_asvspoof_2019}.  However, the term ``audio
deepfake'' evolved to aim to fool humans.  This evolution spurred the
ASVspoof2021 DeepFake~\cite{yamagishi_asvspoof_2021} and Audio Deep Synthesis Detection~\cite{yi2022add}  challenges. Current detection methods primarily employ complex Neural Networks~\cite{
wang2020deepsonar,wang2021investigating, wijethunga_deepfake_2020,jiang2020self,subramani2020learning,zhang2021fake, khalid2021evaluation,tak2022automatic,martin2022vicomtech}.  These models generally focus on low-level features (e.g., spectrogram, MFCC, and CQCC).

Recent work has explored the comparison between the abilities for humans to act as deepfake detectors~\cite{muller2022,warning_humans,wenger2021,allyour}. While humans do not perform as well as most detection models, Warren et al.~\cite{WTC_24} demonstrate that models do not strictly improve upon human performance, but rather have a difference in the way that they detect, and that both are necessary in the detection process. They demonstrate that humans are more sensitive to false negative decisions (i.e., believing deepfakes are humans), while models are sensitive to false positive decisions (i.e., believing humans are deepfakes). Several of these studies~\cite{WTC_24,warning_humans} show that humans rely on linguistic features like prosody, pace, disfluencies and accents to aid in their decision process. Our work aims to explore the ability for models to detect using the same features as humans. 

\section{Prosodic Analysis} \label{sec:acoustic}

\subsection{Prosody Elements} \label{sec:elements}
Structured language is defined not just by the words we use, but also by the acoustic features of
our voices, known as prosody. 
Prosody is a linguistic catch-all for the
suprasegmental parts of speech (e.g., stress, pitch, and tone). In linguistics, prosodic analysis is generally used to capture
abnormal variations in prosody, voice quality, and pronunciations to diagnose
pathological speech~\cite{AcousticPhonetics,diehl_watson_bennetto_mcdonough_gunlogson_2009}.   For connected
speech (i.e., conversations), we analyze prosody through a prosodic feature set and a voice quality
feature set.

The prosodic features that are most consistently used in linguistics are pitch,
length (i.e., duration of syllable sounds), and loudness. For 
this paper, we focus on the prosodic feature of pitch and its
related concept of intonation which helps describe the complexity of human
speech. Aspects of 'pitch' and 'tone' are the prosodic features that humans focus on for detection~\cite{WTC_24}. Voice quality is represented by another
set of features that classify the raspy and airy nature of human voices. The voice quality
measurements that are used within acoustic studies are jitter, shimmer, and
harmonics-to-noise ratio (HNR). 
Some studies consider jitter and shimmer to be
prosody-based instead of voice quality depending on their use in either voice sounds or 
speech sounds. 
All of these features (e.g., pitch, jitter, shimmer, and HNR) will serve as the basis of our prosodic analysis and will help determine vocal abnormalities for the detection of synthetically generated speech. 

\subsubsection{Fundamental Frequency and Pitch} 
Fundamental frequency ($F_0$), the acoustic measurement of pitch, is a basic
feature that describes human speech. Frequency is the number of times a sound
wave repeats during a given period and the fundamental frequency is the lowest
frequency of a voice signal~\cite{vocal_analysis}.  Similarly, pitch is defined
as our brain's perception of the fundamental frequency. The difference in the
two features is most apparent when it comes to phantom fundamentals. A phantom
fundamental is a phenomenon in which the brain perceives harmonics of the
fundamental frequency as the existence of the fundamental frequency even if it is
missing or removed from the voice signal (e.g., spectral filtering and
frequency modulation). The fundamental frequency for male speakers is typically
around 125 Hz, while female speakers tend to average around 210
Hz~\cite{frequency_eriksson}.  Whenever the fundamental frequency is present,
the fundamental frequency and pitch refer to the same value. 

\subsubsection{Intonation}
The rise and fall of a person's voice (i.e., melodic patterns) refer to the
prosodic feature called intonation. Varying tones help to give meaning to an utterance,
allowing a person to stress certain parts of speech and express the desired
emotion. A classic example of this is in the English language where one can
turn any statement into a question by adding a rising intonation at the trail
of the sentence. Without emotion, conversations become lifeless and unengaging,
which is why varying intonation helps us indicate liveliness and makes speech
sound more natural. A shift from a rising tone to a falling one is known as
peaking intonation and the inverse is called dipping intonation. The more
frequently these appear in speech, the less monotone a person will sound.

\subsubsection{Jitter and Shimmer}
Voiced speech comes from a fluctuating organic source, making it quasi-periodic,
and creating measurable differences in the oscillation of the signal.  Jitter is
the frequency variation between two cycles (i.e., period length), and shimmer
measures the amplitude variation of a sound wave. Jitter comes from lapses in
control of our vocal cord vibrations and people diagnosed with speech pathologies generally have higher amounts of jitter in their voice. The jitter levels in a person's voice are a
representation of how ``hoarse'' their voice sounds~\cite{vocal_measures}.
Shimmer, however, corresponds to the presence of breathiness or noise emissions
in our speech~\cite{vocal_analysis}.  Both of these features capture the subtle
inconsistencies that are present in human speech.

\subsubsection{Harmonic to Noise Ratio (HNR)}
Harmonic to noise ratio is the ratio of periodic and non-periodic components
within a segment of voiced speech~\cite{hnr_murphy}. The HNR of a speech sample
is commonly referred to as harmonicity and measures the efficiency of a
person's speech. With respect to prosody, HNR denotes the texture (i.e.,
softness or roughness) of a person's sound. The combination of jitter, shimmer,
and HNR quantifies an individual's voice quality.  Harmonicity is another
measurement that can help determine speech pathology or asthenia (i.e.,
abnormal physical weakness) in the voice.  Most of the sounds that are made
during human speech are associated with high HNR values.  People, however, have
varying degrees of lung capacity and strength in their vocal cords which makes
HNR dependent on each individual, even for the same sound. 

\subsection{Acoustic Prosodic Analysis}

The prosodic features presented in Section~\ref{sec:elements} are the main elements that constitute traditional prosodic
analysis, which has appeared in linguistics for decades~\cite{voice_quality}. 
Acoustic prosodic analysis uses combinations of these prosody features to diagnose speech pathologies and to better understand the limitation in expressiveness and deficits in communications for people with a variety of behavioral and intellectual issues. This analysis looks for abnormalities and deviations in prosodic features from the expected range to determine issues with an individual's voice and compares those results with various known speech pathology deviations. By treating deepfakes as speech pathologies or disorders, we can leverage the use of acoustic prosodic analysis in deepfake detection to differentiate real and fake speech. Using various combinations of these features helps to describe the complex patterns and properties of an individual's voice. 

\begin{figure}
    \begin{center}
        \includegraphics[width=\columnwidth]{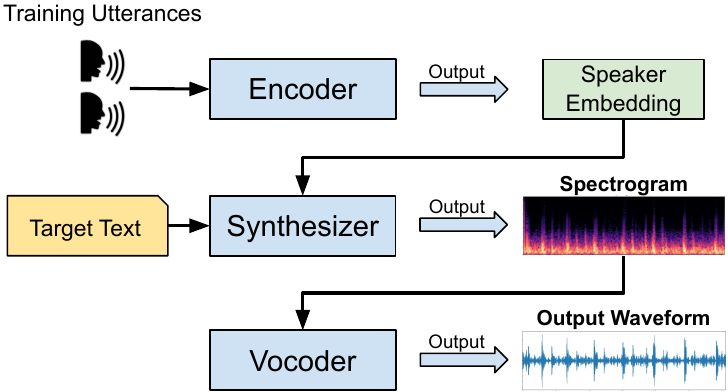}
        \caption{The standard generation for audio deepfakes is an
        encoder/synthesizer/vocoder pipeline. The encoder
        generates a voice embedding, the synthesizer 
        creates a spectrogram for a target phrase, and the vocoder converts
        the spectrogram into a synthetic waveform.} 
        \label{fig:deepfake_pipeline}
    \end{center}
\end{figure}

\section{Taxonomy of Fake Speech} \label{sec:taxonomy}

The term ``audio deepfake'' can refer to different types of adversarial
audio. To clarify the multitude of threats, we introduce a taxonomy to
characterize deepfake attacks by their objective, generation method,
and intended recipient.

\subsection{Objective} \label{taxonomy_objective}

While deepfakes can be created for many purposes, the objective of the attack affects both its threat level and impact. While training and generation generally require specific voices, the target of a deepfake can either be claiming to impersonate a specific person or to generate a generic untargeted voice. 

\paragraph{Targeted Deepfake} \label{Targeted Individual}

Targeted deepfakes use Deep Neural Networks (DNNs) to create audio based on a specified
individual. These systems follow a standard encoder, synthesizer,
and vocoder model seen in Figure~\ref{fig:deepfake_pipeline}. 
The encoder is designed to model a speaker's voice.
It uses a set of utterances to create a distinctive
profile of a speaker's voice called a speaker embedding. The more samples 
given to the encoder, the more accurate the embedding will be, with
diminishing returns. Using the speaker embedding 
and an input text, the synthesizer generates a Mel spectrogram for the given
text. The Mel spectrogram uses frequencies converted into the Mel scale, which
is a logarithmic scale designed to mimic the human ear's perception of sound.
Some recent synthesizers can generate the spectrogram from just a character
sequence or phonetic spellings~\cite{tacotron}. 
The vocoder converts the Mel spectrogram 
into the corresponding audio waveform. 
While the synthetic speech generation process is constantly changing with new
tools, these three components are fundamental to the framework. 
The quality of
the models in each tool is dependent on the quantity of training data and
the complexity of the structure. 
The trade-off for better models is an increase
in time and training resources. 

Using these models, an adversary can create a sample for malicious purposes
(e.g., a targeted audio deepfake). For example, a
person may 
try to get a target to transfer
money on fake orders made in the voice of the CEO~\cite{theft}.  
Targeted audio has also been used for benign purposes 
in the medical field (e.g., recreating
the voice of someone who has lost the ability to speak~\cite{MBP14}) 
and is
emerging as a possible future alternative in cinematography (e.g., recreating
audio of deceased actors/actresses or developing movies with completely digital
versions of actors/actresses).

\paragraph{Untargeted Deepfake} \label{Untargeted Individual}

Untargeted deepfakes also use neural networks, typically a single vocoder, to
create fake organic-sounding audio. 
While untargeted audio can also be based on an individual's voice, 
it differs in that the sample is not
claiming to be the person that it was trained on. For example, the voice of
Siri is based on a voice actor, but the assistant uses the sample to sound
more human, not impersonate the actor~\cite{siri_bennett}.
This aims at giving
consumers the feeling of having a real conversation and \emph{not}
that they are talking to a real person.

Adversarially, untargeted attacks can be used for spam in scenarios where the
identity of the speaker does not need to be verified to elicit a response. For
example, 
police dispatch often react to
incoming calls without verifying the caller if the situation presents
itself as dire. 

\subsection{Generation Methodology} \label{gen_method}

\paragraph{Fully Generative Audio} \label{Generative Audio}

Fully generative audio is a classification of fake speech that takes a
representation of a person's voice and creates a speech sample from scratch.
This process comes from recent advancements in machine learning (ML) and DNNs.  
Hidden layers in DNNs allow complicated tasks
such as speech synthesis to work, but have difficulty 
learning specific attributes. 

If speech
synthesis needs to learn specific features such as prosody, the process is
mainly trial and error of training data,
hyperparameters, and activation functions. 

\begin{figure}
    \begin{center}
        \includegraphics[width=\columnwidth,keepaspectratio]{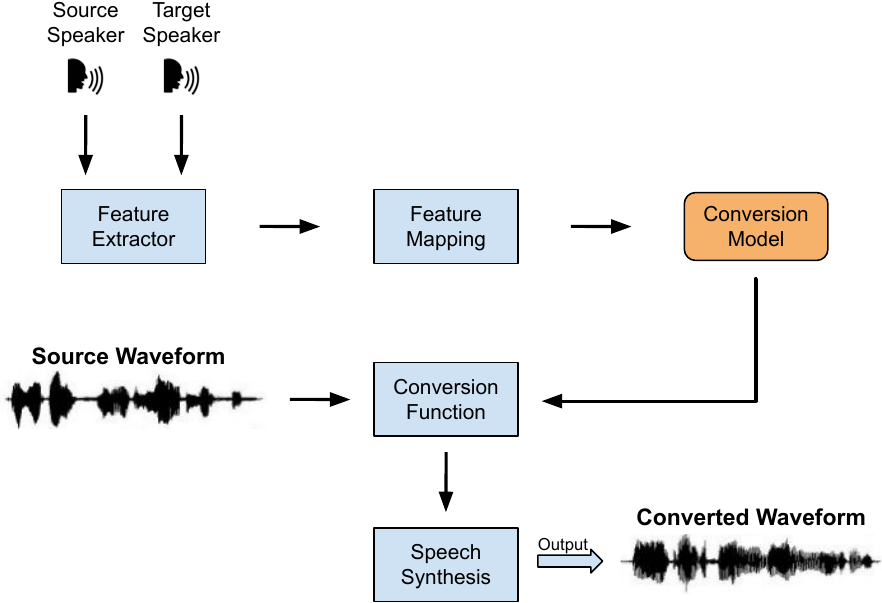}
        \caption{A voice conversion pipeline for a two-speaker setup. 
        A model is trained on the features of both the target and source speakers.
        Using the trained model, the technique takes in a source waveform, transforms it,
        and then synthesizes the target waveform.} 
        \label{fig:VC_pipeline}
    \end{center}
\end{figure}

\paragraph{Voice Conversion}

Voice conversion is a technique that is distinguished by its use of samples
from two speakers: a source speaker and a target speaker as shown in Figure~\ref{fig:VC_pipeline}. Voice conversion uses a transfer function to
convert the spectral features of the source speaker to closely match those of the
target speaker. 
Modern VC systems will typically use the
converted spectral features with a vocoder to generate the final fake speech
sample. 
Generally, voice conversion outputs are noisier than targeted deepfakes
since the system has to account for the background noise in the source sample.
Since this generation method requires a source speaker, it is only a targeted objective.

\subsection{Intended Recipient}
Deepfake samples are aimed at fooling either a machine or a human. The output 
requirements change depending on the target. Thus, designating the intended recipient is
crucial for allocating appropriate tasks.

\paragraph{Machine Recipient}
Audio deepfakes were originally focused, in practice, on targeting machine systems such as 
automatic speaker verification (ASV) systems. This focus was driven by the
low audio quality of early deepfakes. Machine learning models were
able to craft samples that were optimized against the models of an ASV program, as these 
samples did not need to succeed in human authentication. 
This means that deepfake models only needed to make audio 
whose quality and imitation were good enough to bypass the ASV. 

\paragraph{Human Recipient}
Over time, improved deepfake audio quality output has shifted the focus of deepfakes from fooling machines to fooling humans. With the addition of better vocoders and natural language processing models, deepfakes have become eerily believable to the average listener. These deepfakes cause social issues and impose uncertainty on the authenticity of videos and audio clips. Unlike machine-targeted deepfakes, these require sufficient quality for a human to believe the source.

Detection tasks in this space focus on forensic applications on questionable outlets for media, social 
media forums, and social engineering attacks. This type of detection has become popularized 
with recent world events such as the war in Ukraine and the growing trend of faking high-ranked political 
officials and business professionals~\cite{ukraine_deepfake}.

\begin{figure*}[!t]
    \begin{center}
        \includegraphics[width=.8\linewidth]{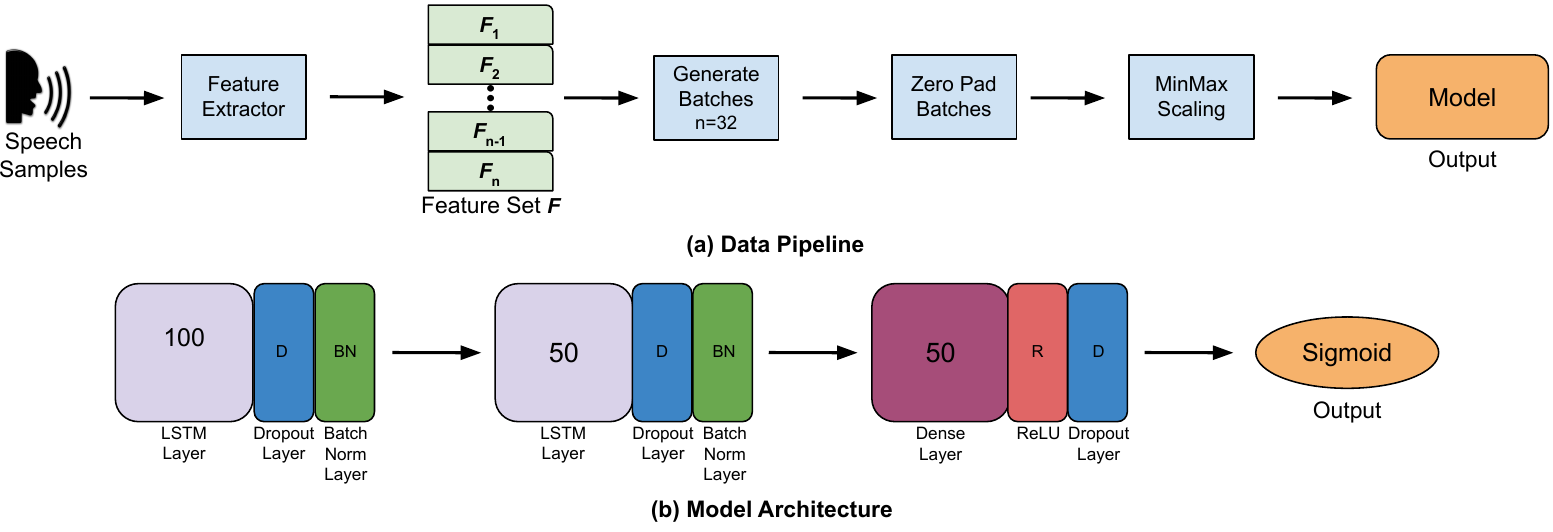}
        \vspace{-0.3cm}
        \caption{ The pipeline for processing speech sample features and the final LSTM model architecture used as our detector.
        Pipeline (a) demonstrates the audio preprocessing steps for feature extraction and batch generation used to train our LSTM architecture.
        The model architecture in (b) shows the size and activation units for each hidden layer. The details to the architecture
        allow the model to be reproduced to duplicate or verify results. Combined, this pipeline and
        model form our classification process.
        } 
        \label{fig:training_pipeline}
    \end{center}  
\end{figure*}
\section{Reserch Questions and Approach} 
\label{sec:rqs} 

Based on the taxonomy, we design our experiments and reserach quesitons around deepfakes designed to fool a person. We aim to answer the following research questions:

\begin{rqs}

    \item \textbf{Viability of Prosody for Detection:} Do prosodic features robustly distinguish between deepfake and human generated audio?
    
    \hfill
    \item \textbf{Benefits of Using A Prosody Detector:} What are the benefits to using a prosody detector approach over the standard deepfake detection model?
    
\end{rqs}

Similar to a user study, we perform our testing from a forensic postprocessing standpoint. The model, like a user, processes the entire audio sample and makes one classification of human or computer generated for the entire clip. Additionally, this approach aims to justify the classification of the detector and explain which features (i.e., prosody elements) impact the decision. 

\section{Dataset} \label{sec:dataset}

For training and evaluating our model, we focus on the widely used ASVspoof2021
dataset~\cite{yamagishi_asvspoof_2021}. This dataset contains samples that are clearly defined 
for our task and represents the community standard used for deepfake detectors. Additionally, ASVspoof2021 provides a set of baseline models tested against the dataset for comparison. 

\subsection{ASVspoof2021:} \label{subsec:meth_dataset_asvspoof} 
The ASVspoof dataset iterations are considered state-of-the-art for each subsequent 
ASVspoof challenge (2015-2023), and as such we aim to use one of the recent dataset releases.
Thus we use the ASVspoof2021 dataset to train and validate our model. ASVspoof2021 contains 
three datasets: training, validation, and evaluation. The ASVspoof2021 dataset contains physical access data, logical access data, and deepfake data. The physical access (PA) data contains spoofing attacks performed at the sensor level, such as replay attacks, and rely on weakness in the automatic speech recognition system hardware. The logical access (LA) data focuses on spoofing attacks generated by text-to-speech (TTS) and voice conversion (VC) aimed at speaker verification. The deepfake audio also focuses on TTS and VC audio with two differences: (1) the use of more generalized compression algorithms and (2) the focus on audio forensics removing the use of ASV systems. The PA/LA datasets are focused more on bypassing speaker verification systems, while the deepfake set only focuses on whether the audio is human generated. Thus, we focus on the deepfake data which contains a collection of 'bonafide' and synthetic audio samples that were processed with various lossy codecs. The source data for the ASVspoof2021 evaluation dataset is a combination of the previous year's logical accessed data along with many other new sources for a significantly larger evaluation set over previous challenge iterations. This results in attack audio generated with more than 100 different audio spoofing algorithms. The training and validation sets contain different samples generated from the same 20 speakers, however, the evaluation set does not distinguish between different speakers for their audio files. The total number of samples in each set are \textbf{training}: 22,800 deepfakes and 2,580 bonafide; \textbf{validation}: 22,296 deepfakes and 2,548 bonafide; and \textbf{evaluation}: 589,212 deepfakes and 22,617 bonafide.  

\section{Prosody Deepfake Detector} \label{sec:model}

We develop a deepfake detection model based on the elements of prosodic analysis as defined in Section~\ref{sec:elements} to explore an alternative approach to deepfake detection systems. Our data pipeline is as follows: extract pitch prosodic features from speech samples, standardize the extracted feature with standard scaling, and then predict with our model, as shown in Figure~\ref{fig:training_pipeline}. The six prosody elements we use are the average and standard deviation of the mean-$F_0$ of the window, jitter, shimmer, and the average and standard deviation of HNR of the window. We perform a hyper-parameter search to find the best parameters for extracting pitch from the audio, the optimal window size to calculate our features over, and the best model architecture.

\subsection{Feature Extraction}

Our feature extraction process collects the common prosodic features which can be measured using speech analysis tools. We extract the prosodic parameters discussed in Section~\ref{sec:acoustic} using Parselmouth~\cite{parselmouth}, a Python interface for the state-of-the-art acoustic analysis tool called Praat~\cite{praat}. Figure~\ref{fig:training_pipeline} shows the feature extractor in the pipeline. For full equations on any of our measured features (e.g., HNR, jitter, and shimmers) refer to the Appendix.

\begin{figure*}
    \begin{center}
        \includegraphics[width=\linewidth]{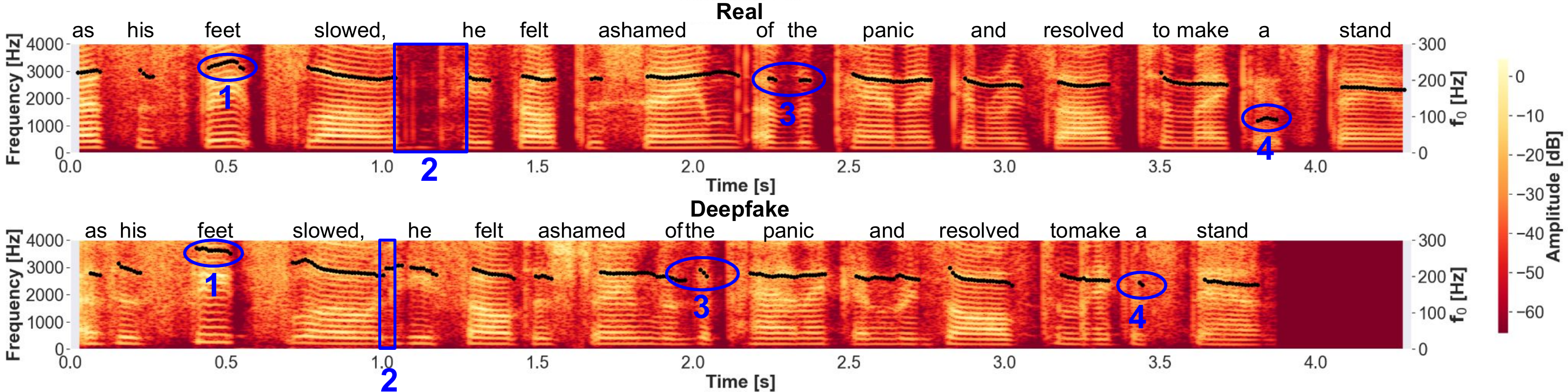}
        \caption{Examples of the spectrogram and fundamental frequency sequences for an organic 
        and synthetic audio sample. The top graph is an organic speaker. The bottom graph is a deepfake trained on the same organic speaker 
        and generated to say the same sentence. 
        Highlighted generation issues illustrate (1) inflection changes, (2) pause discrepancies, and 
        (3,4) combinations of inflection changes, pause discrepancies, and pitch variance.} 
        \label{fig:two_spectrograms}
    \end{center}
\end{figure*}

\subsubsection{Measured Prosodic Features} 
We start by collecting the mean fundamental frequency (mean $F_0$) and standard deviation of $F_0$. To determine the pitch range (i.e., minimum allowed $F_0$ and the maximum allowed $F_0$) required for analysis, we refer to the standard values set to 75 Hz for min $F_0$ and 500 Hz for max $F_0$. We do not adjust these values per audio file and assume that all voiced speech in any given sample falls within this range. 

The fundamental frequency sequence used to get the mean $F_0$ is a series of $F_0$ values sampled with respect to time. These $F_0$ values are shown in Figure~\ref{fig:two_spectrograms} as the dots that make up the black lines. The $F_0$ sequences of human and deepfake speech are similar, but even for the same sentence and speaker they are not the same. Intonation is the changes to the $F_0$ sequence over the length of the audio sample. The differences in $F_0$ sequence are demonstrated in Figure~\ref{fig:two_spectrograms} with the fake audio sample being shorter and words such as ``he'' where the organic audio has a dipping intonation versus peaking intonation in the fake. These distinctions demonstrate that synthetic audio generates pitch without perfectly mimicking the correct $F_0$ sequence. 

We also collect measurements of jitter and shimmer for each sample. We focus on only collecting the values for local jitter and local shimmer as these are commonly used when determining voice quality. The local version of jitter and shimmer look exclusively at consecutive intervals or periods and determines the average difference between them.

The last feature extracted is the harmonicity (HNR). We collect both the mean value and standard deviation of the HNR throughout the audio sample. We assume a standard minimum pitch value of 75 Hz and one period per window to avoid sensitivity to dynamic change in the signal.

\subsubsection{Data Scaling}
After we extract the features, we preprocess the data by standardizing 
the data with Min-Max scaling. Standardizing the data ensures that no feature influences the model 
more than another strictly due to its magnitude. We standardize by scaling the values between 0 and 1 
by taking the difference of the feature value from the minimum and diving that by the range of the feature values.
Formally, given a data matrix $X$, we scale each feature column, $x$, such that for any $x_i$ in
$x$, 

\begin{equation}
    x_{i}^{\text{scaled}}=\frac{x_i - \text{min}(x)}{\text{max}(x) - \text{min}(x)}.
\end{equation} 

\subsubsection{Praat Parameter Search}
While the tools used in applied linguistics are good for diagnosing speech problems, the standard parameters used by the tools are not suited for deepfake detection because they assume an organic source. Some of the abnormalities that can exist in generated speech are ignored or dropped by these tools since they could not be naturally produced by humans. Due to this, we do not use the default parameters when calculating the $F_0$ sequences for the pitch objects. Instead we perform a parameter search to explore the feature space and determine the best combination of values to use in calculating the pitch. 

We focus our search around four parameters: silence threshold, octave cost, octave jump cost, and voiced/unvoiced cost. These are four of the input parameters for determining $F_0$ and the parameters that have the closest relation to speech. The silence threshold gives the amplitude bound for what is determined as silence. The octave cost determines how much high-frequency values for $F_0$ are favored in the case of perfectly periodic signals, while the octave jump cost represents the amount of large frequency jumps that should not be ignored in the signal. The voiced/unvoiced cost figures 
out the sensitivity to transitions between voiced and unvoiced signals. For each of these features, we collect data using a time-series analysis by windowing the audio sample. When training our detector, we test window sizes at 50, 100, 200,and 500 ms non-overlapping window frames. 

To determine general bounds for the parameters, we start our search by processing all of our training data using a grid search for each of the four parameters. We determine which parameter values between zero and one returned a large number of NaN values, signifying inappropriate values for one or more parameters. After determining the bounds, we ran a randomized search of values within those bounds and trained models using an LSTM architecture with three layers (LSTM-64 nodes, LSTM-32 nodes, and Dense-32 nodes). We trained ~2,200 models
for 200 epochs under an ADAM optimizer with a learning rate of 0.0001, and determined the best parameters optimized over the Equal Error Rate (EER). We choose to use EER to allow direct comparison between our system and existing detectors. 

\subsection{Final Architecture Selection \& Training}
Using the optimal value combination for feature extraction, we test five variations on the model architecture. We modify the number of layers and the number of nodes in each layer, but maintain a sigmoid output layer. For an outline of the five model architectures we test, please refer to the Appendix.
The five models we consider are:

\begin{figure}
    \begin{center}
    \includegraphics[width=\columnwidth]{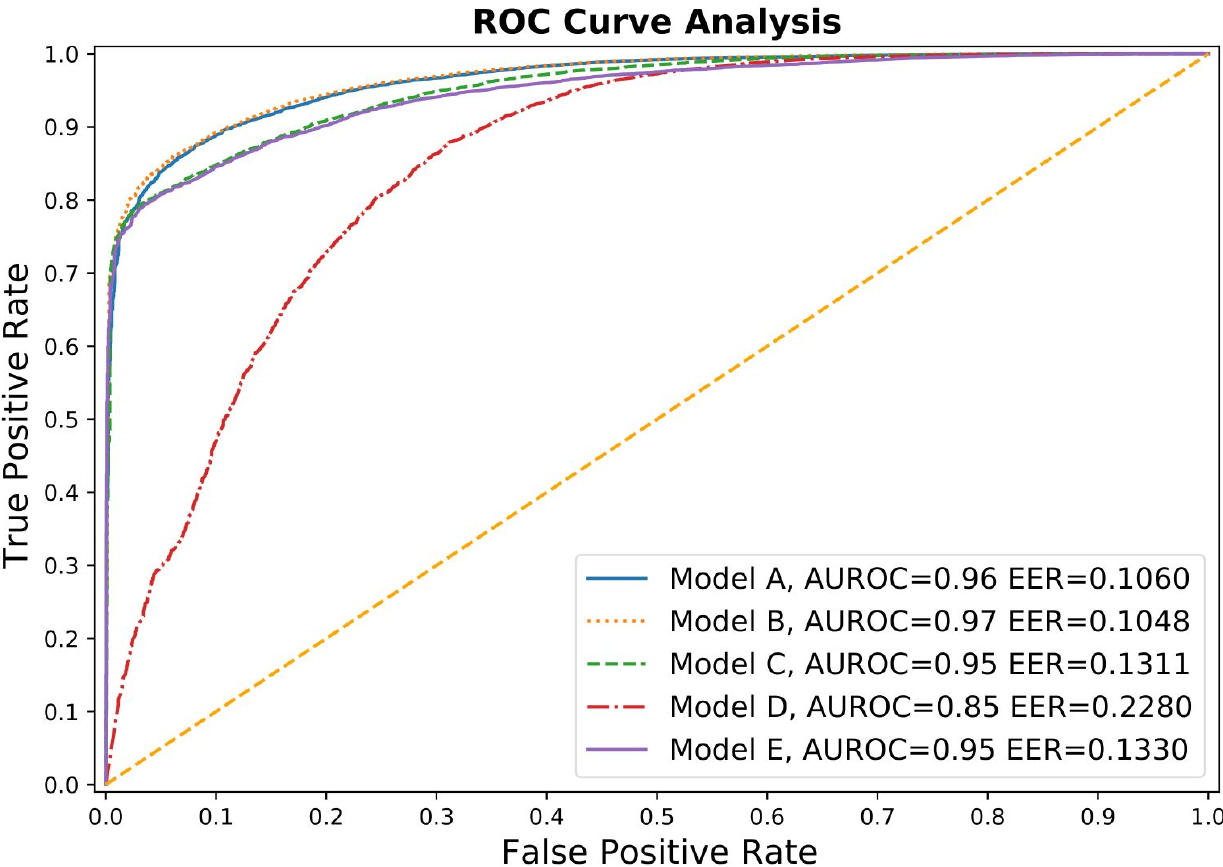}
    \caption{AUROC/EER for each model architecture tested against the
    ASVspoof2021 validation dataset.}
    \label{fig:model_selection}
    \end{center}
\end{figure}

\hfill
\begin{itemize}
    \item {Model A:} Three layers (LSTM with 64 nodes, LSTM with 32 nodes, and Dense with 32 nodes).
    \item {Model B:} Three layers (LSTM with 100 nodes, LSTM with 50 nodes, and Dense with 50 nodes).
    \item {Model C:} Five layers (LSTM layers with 64, 32, 16, 8 nodes, and Dense with 8 nodes).
    \item {Model D:} Two layers (LSTM with 100 nodes, and Dense with 100 nodes).
    \item {Model E:} Three layers (LSTM with 16 nodes, LSTM with 16 nodes, and Dense with 16 nodes).
\end{itemize}

\hfill

We train each model for 200 epochs with an ADAM optimizer that has a learning rate of 0.0001. The performance of each model can be seen in Figure~\ref{fig:model_selection}. Our best model (Model B) is optimized for performance by a hyperparameter search over the pitch calculation parameters and model architecture.

We then train our best model using the ASVspoof2021 training data. We perform a frame-level feature extraction where we give voiced frames the extracted prosodic feature values and unvoiced frames a zero value. Each sample within a batch is also zero padded with unvoiced frames to the longest sample length, which is the same methodology as the ASVspoof2021 baselines. 

\subsection{Final Model Evaluation}\label{sec:trans}

To determine the overall performance of our prosody model, we process the ASVspoof2021 deepfake evaluation dataset using the same data processing methodology we train with.
To show our model's performance, we use the Equal Error Rate. The EER is defined as the model's probability threshold at which the false negatives are equal to the false positives. For this metric, an EER closer to zero is indicative of a stronger model. The ASVspoof2021 competition continues to use this metric despite its deprecation to have a common metric between all challenges. 

The ASVspoof2021 challenge is an anonymous submission competition. The only required metric for reporting is EER. Submissions to the challenge are not required to publish documentation. This makes verifying and reproducing the top results infeasible. Because of the requirements of the 
competition, these models also do not report standard performance metrics such as precision, recall, or $F_1$-score. The ASVspoof2021 challenge does not require adversarial testing on the submitted detectors and does not give robustness guarantees on their performance. The challenge does provide access to four baseline models which are used as benchmarks for the competitors. Using these baselines, we can recreate those models in order to test against a larger set of performance metrics for comparing results.

\begin{table*}[]
    \centering
    \caption{Results of the training, validation, and evaluation experiments of our best model (Model B) on ASVspoof2021. Each experiment represents the classification of audio in the ASVspoof2021 deepfake track training, validation, and evaluation datasets respectively. Baselines 01-04 are the baseline models that were provided for comparison in the ASVspoof2021 challenge. Due to reporting rules of the competition, the standard performance metrics are not reported and the only metric available for each baseline is the EER. Due to this limitation, we report the measured performance metrics for each model after reproducing each baseline model.}
    \label{tbl:models}
        \resizebox{\textwidth}{!}
        {\begin{tabular}{|c|ccc|cc|cc|cc|cc|}
        \hline
        \multicolumn{1}{|l|}{}                   & \multicolumn{3}{c|}{Our Model}                                                                                                               & \multicolumn{2}{l|}{Baseline-01 (LFCC-GMM)}                                         & \multicolumn{2}{l|}{Baseline-02 (CQCC-GMM)}                                         & \multicolumn{2}{l|}{Baseline-03 (LFCC-LCNN)}                                        & \multicolumn{2}{l|}{Baseline-04 (RawNet)}                                           \\ \cline{2-12} 
        \multicolumn{1}{|l|}{}                   & \multicolumn{1}{l|}{Training}                       & \multicolumn{1}{l|}{Validation}                      & \multicolumn{1}{l|}{Evaluation} & \multicolumn{2}{c|}{Evaluation}                                                     & \multicolumn{2}{c|}{Evaluation}                                                     & \multicolumn{2}{c|}{Evaluation}                                                     & \multicolumn{2}{c|}{Evaluation}                                                     \\ \cline{2-12} 
        \multicolumn{1}{|l|}{\multirow{-3}{*}{}} & \multicolumn{3}{c|}{Measured}                                                                                                                & \multicolumn{1}{c|}{Reported}                        & Measured                      & \multicolumn{1}{c|}{Reported}                        & Measured                      & \multicolumn{1}{c|}{Reported}                        & Measured                      & \multicolumn{1}{c|}{Reported}                        & Measured                      \\ \hline
        Accuracy                                 & \multicolumn{1}{c|}{\cellcolor[HTML]{FFFFFF}97\%}   & \multicolumn{1}{c|}{\cellcolor[HTML]{FFFFFF}94\%}    & \cellcolor[HTML]{FFFFFF}93\%    & \multicolumn{1}{c|}{\cellcolor[HTML]{FFFFFF}-}       & \cellcolor[HTML]{FFFFFF}47\%    & \multicolumn{1}{c|}{\cellcolor[HTML]{FFFFFF}-}       & \cellcolor[HTML]{FFFFFF}46\%    & \multicolumn{1}{c|}{\cellcolor[HTML]{FFFFFF}-}       & \cellcolor[HTML]{FFFFFF}93\%    & \multicolumn{1}{c|}{\cellcolor[HTML]{FFFFFF}-}       & \cellcolor[HTML]{FFFFFF}95\%    \\ \cline{1-1}
        Precision                                & \multicolumn{1}{c|}{\cellcolor[HTML]{FFFFFF}0.91}   & \multicolumn{1}{c|}{\cellcolor[HTML]{FFFFFF}0.86}    & \cellcolor[HTML]{FFFFFF}0.61    & \multicolumn{1}{c|}{\cellcolor[HTML]{FFFFFF}-}       & \cellcolor[HTML]{FFFFFF}0.53 & \multicolumn{1}{c|}{\cellcolor[HTML]{FFFFFF}-}       & \cellcolor[HTML]{FFFFFF}0.66 & \multicolumn{1}{c|}{\cellcolor[HTML]{FFFFFF}-}       & \cellcolor[HTML]{FFFFFF}0.64 & \multicolumn{1}{c|}{\cellcolor[HTML]{FFFFFF}-}       & \cellcolor[HTML]{FFFFFF}0.68 \\ \cline{1-1}
        Recall                                   & \multicolumn{1}{c|}{\cellcolor[HTML]{FFFFFF}0.94}   & \multicolumn{1}{c|}{\cellcolor[HTML]{FFFFFF}0.82}    & \cellcolor[HTML]{FFFFFF}0.66    & \multicolumn{1}{c|}{\cellcolor[HTML]{FFFFFF}-}       & \cellcolor[HTML]{FFFFFF}0.68 & \multicolumn{1}{c|}{\cellcolor[HTML]{FFFFFF}-}       & \cellcolor[HTML]{FFFFFF}0.52 & \multicolumn{1}{c|}{\cellcolor[HTML]{FFFFFF}-}       & \cellcolor[HTML]{FFFFFF}0.77 & \multicolumn{1}{c|}{\cellcolor[HTML]{FFFFFF}-}       & \cellcolor[HTML]{FFFFFF}0.8  \\ \cline{1-1}
        F1 Score                                 & \multicolumn{1}{c|}{\cellcolor[HTML]{FFFFFF}0.92}   & \multicolumn{1}{c|}{\cellcolor[HTML]{FFFFFF}0.84}    & \cellcolor[HTML]{FFFFFF}0.63    & \multicolumn{1}{c|}{\cellcolor[HTML]{FFFFFF}-}       & \cellcolor[HTML]{FFFFFF}0.37 & \multicolumn{1}{c|}{\cellcolor[HTML]{FFFFFF}-}       & \cellcolor[HTML]{FFFFFF}0.36 & \multicolumn{1}{c|}{\cellcolor[HTML]{FFFFFF}-}       & \cellcolor[HTML]{FFFFFF}0.68 & \multicolumn{1}{c|}{\cellcolor[HTML]{FFFFFF}-}       & \cellcolor[HTML]{FFFFFF}0.72 \\ \cline{1-1}
        \rowcolor[HTML]{D9D9D9} 
        EER                                      & \multicolumn{1}{c|}{\cellcolor[HTML]{D9D9D9}3.99\%} & \multicolumn{1}{c|}{\cellcolor[HTML]{D9D9D9}10.48\%} & 24.72\%                         & \multicolumn{1}{c|}{\cellcolor[HTML]{D9D9D9}25.56\%} & 25.50\%                      & \multicolumn{1}{c|}{\cellcolor[HTML]{D9D9D9}25.25\%} & 25.40\%                      & \multicolumn{1}{c|}{\cellcolor[HTML]{D9D9D9}23.38\%} & 22.90\%                      & \multicolumn{1}{c|}{\cellcolor[HTML]{D9D9D9}22.38\%} & 22.10\%                      \\ \hline
        \end{tabular}}
\end{table*}

As previously stated, ASVspoof2021 provides four baseline models.
Nautsch et al.~\cite{nautsch2021asvspoof} uses a Gaussian Mixture Model (GMM) to
detect deepfakes. They implement an LFCC method and a CQCC model for Baseline-01 and Baseline-02, respectively. Baseline-03 by Wang et al.~\cite{wang2021comparative} constructs a Light Convolutional Network (LCNN) which operates on an MFCC. The best model is Baseline-04, a RawNet2 model~\cite{9414234}. The baseline models 01-04 achieve EERs of 25.56\%, 25.25\%, 23.48\%, and 22.38\%, respectively~\cite{yamagishi_asvspoof_2021}. Overall, our 24.7\% EER is directly in the middle of the four baselines. 

Several prior studies have demonstrated that single metrics, especially EER, do not give the complete representation of a model's performance and inherently hide the system's trade-offs when implemented~\cite{needlestack_layton,osti_10091768}. To better compare and understand our results, we train versions of the baseline models and calculate additional performance metrics for comparison such as accuracy, percision, recall, and $F_1$-score. These metrics are provided in addition to the EER in Table~\ref{tbl:models}. We use the reported EER values from the competition to validate that our trained models perform the same as the ones used in the competition. As seen in Table~\ref{tbl:models}, our prosody-based model also performs as well as the baseline models in all of the other standard machine learning performance metrics.
\\
\\ \noindent\fbox{\parbox{\columnwidth}{\textbf{Finding 1.} \emph{A prosody model performs comparatively with other models when detecting deepfakes across all standard machine learning model performance metrics. (\textbf{RQ1})}}}
\\ 

\section{Adversarial Testing and Explainability} \label{sec:adversarial}

While the prosody model has a similar performance to other detetors, we explore some of the benefits that a prosody-based approach has over other detectors. Some limitations to standard deep learning models are their black-box nature that minimizes explainability and increases their susceptibilty to adversarial optmization attacks. In this section, we demonstrate the explainability of individual influential features on the model decisions and test our model under adversarial conditions.

\begin{figure}
	\centering
	\includegraphics[width=\linewidth]{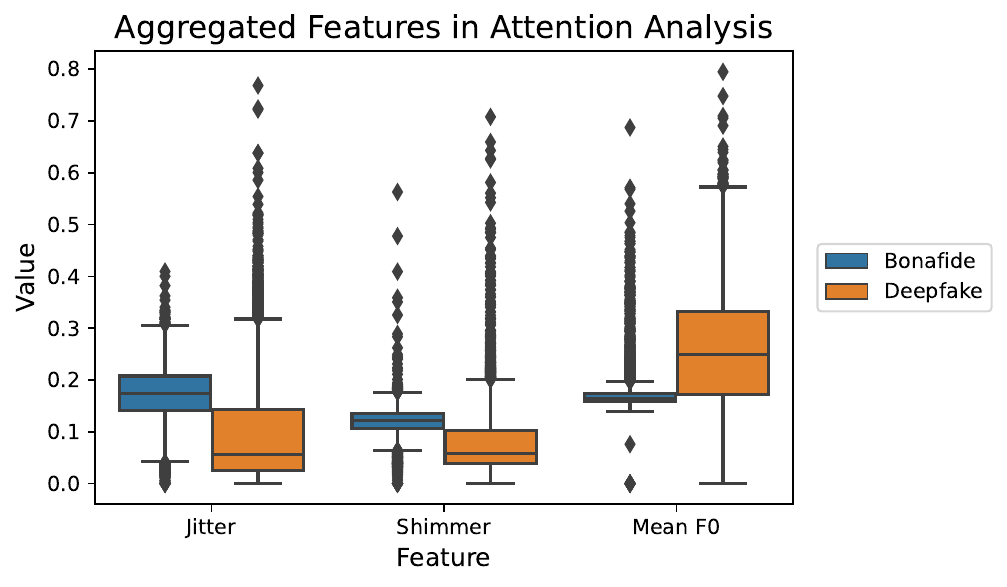}
	\caption{For each sample in the eval dataset, we compute its attention
	vector to find what time-slice most affected the classification. Then, 
	we show the aggregated information for Jitter, Shimmer, and Mean $F_0$.}
	\label{fig:attention}
\end{figure}

\subsection{Attention}
We implement attention mechanisms to explain 
what features are most affecting the classification.
Attention mechanisms, first proposed by Bahdanau et al.~\cite{bahdanau2014neural},
seek so as to allow networks broader access to better infer long-term effects
in a sequence. These attention mechanisms specifically provide
hidden states of a sequence to all of the subsequent layers.
One can use the attention vector to weigh the importance
of each input sequence. Weng et al.~\cite{weng2018attention} provide a
thorough analysis of this attention mechanism and its variants. We integrate
Self-Attention by Cheng et al. which compares
the input sequence against itself to understand which
part of the input sequence greatly influenced the prediction~\cite{cheng2016long}.

In our case, we can retrain our model with a Self-Attention layer after our input layer. 
For each sample in the development set, we 
compute the attention vector which tells us which time-slice affected the classification the most. This
time-slice consists of the input prosodic features calculated in that
interval. Because we trained on prodosic features, we can isolate the differences in feature values that occur during 
the interval. We visualize the differences for the Jitter, Shimmer, and
Mean $F_0$ in the most important time-slice, as calculated by the
attention mechanism, between the bonafide and deepfake samples
in Figure~\ref{fig:attention}. We see that for Jitter and Shimmer,
the bonafide samples on average have smaller values. For 
Mean $F_0$, the deepfake samples have smaller values. 
\\
\\ \noindent\fbox{\parbox{\columnwidth}{\textbf{Finding 2.} \emph{Unlike standard black-box machine learning models, we are able to use attention mechanisms to determine which prosodic elements influenced the detection decision. This adds a layer of explainability to our model that is missing from other detectors. (\textbf{RQ2})}}}
\\

\subsection{Adversarial Experiments}
While we have explored the general performance of the prosody model, we also need to test the robustness of the model. Previous attempts at audio deepfake
detection do not consider the robustness to adversarial samples
against their model. We use the best performing baseline model, RawNet2, 
to demonstrate the brittle nature of their results in the presense of an
adaptive adversary.

The Iterative Least-Likely Method~\cite{kurakin2016adversarial}
takes the gradient of the loss function with respect
to the input vector and successively applies perturbations in the direction of the gradient.
Formally, given an input $X$ with label $y_{true}$, we can
construct $X^{adv}$ that has the target label $y_{t}$\footnote{For a binary classification problem, $y_{t}$ is the Least-Likely Class.} by iteratively applying to $X_{0}^{adv}= X$:
\begin{equation*}
\begin{matrix}
  X^{adv}_{N+1} = Clip_{X,\epsilon}\{X^{adv}_{N} + \alpha\text{sign}(\nabla_X J(X^{adv}_{N}, y_{t})) \},
\end{matrix}
\end{equation*}
where $Clip$ constrains the perturbation to an $L_{\infty}$ ball
around $X$ with radius $\epsilon$, $\alpha$ is the magnitude
of the perturbation, and $\nabla_X$ is the gradient of the loss function $J$
with respect to the input.

Formally, the perturbation, $\delta$, required to
turn a benign feature vector, $x$, with label, $y_{0}$, into an adversarial sample
is defined as, 

\begin{equation*}
\eta = \epsilon\cdot \text{sign}(\nabla_xL(\theta,x,y)),
\end{equation*}
where $\epsilon$ is the magnitude of the perturbation,
$\nabla_x$ is the gradient of the loss function, $L$, with
model parameters $\theta$.
FGSM 
\begin{figure}[!t]
	\begin{center}
			\includegraphics[width=\linewidth]{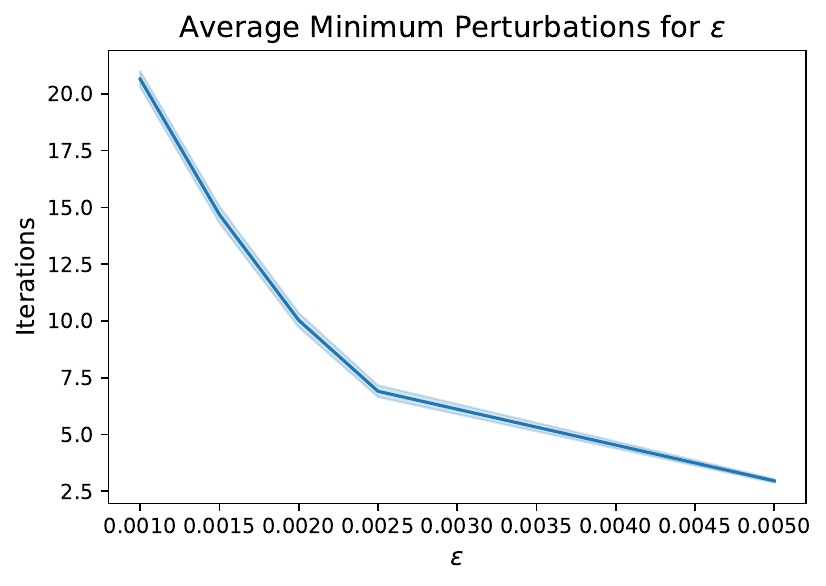}
			\vspace{-0.3cm}
			\caption{ 
				The average number of steps to successfully craft the adversarial sample with label $y_{t}$ for each $\epsilon$. 
				We see that the average minimum perturbations converges as we let $\epsilon$ get larger. This shows that
				our adversarial attacks are convergent.
				For $\epsilon=0.001$, it takes significantly more steps (average 22 steps), but for the larger $\epsilon$s,
				we see that it takes less than 10 steps to craft an adversarial sample. With $\epsilon=0.005$, the average 
				needed perturbations is only 3 steps.
			} 
			\label{fig:min_pert}
		\end{center}
	\end{figure}

\noindent\textbf{Experiments:} We use the pitch calculations as defined by Boersma~\cite{boersma1993accurate}.
The pitch calculations  use a non-differentiable, path-finding algorithm.
This inhibits any gradient-based attacks, and thus, \emph{our system cannot be successfully attacked by a gradient based adaptive adversary}.
Conversely, the baseline models are fully differentiable, thus we can create a simple gradient-based adaptive adversary for their models.

To create the adversarial sample, we let $X_{i,j}$ be the matrix of the audio file data with $i$ representing the time domain and $j$ representing the number of channels of the audio file. We choose $\epsilon$ to minimize the overall perturbance of the audio sample (Figure~\ref{fig:min_pert}).
We choose
$\epsilon \in \{ 0.001, 0.0015,0.002, 0.0025, 0.005 \}$, as 
$X_{i,j} \in [-1,1] \forall i,j$. Thus, each 
$\epsilon$ is perturbing less than $0.5\%$ of each data point. We set $\alpha$ to 0.001. 
Essentially, we are adding small amounts of white noise to the
samples. 

We sample 2800 audio samples from the ASVspoof2021 evaluation data set. Then, we
craft adversarial samples for the Baseline-RawNet2 for each sample using the five different $\epsilon$s. 
Even with the largest amount of added white noise, we still maintain an average waveform amplitude distribution analysis signal to noise ratio (WADA-SNR)~\cite{Kim2008RobustSR} $>41dB$,
which is considered excellent quality audio~\cite{snr_quality}.

\begin{figure}[!t]
	\begin{center}
			\includegraphics[width=\linewidth]{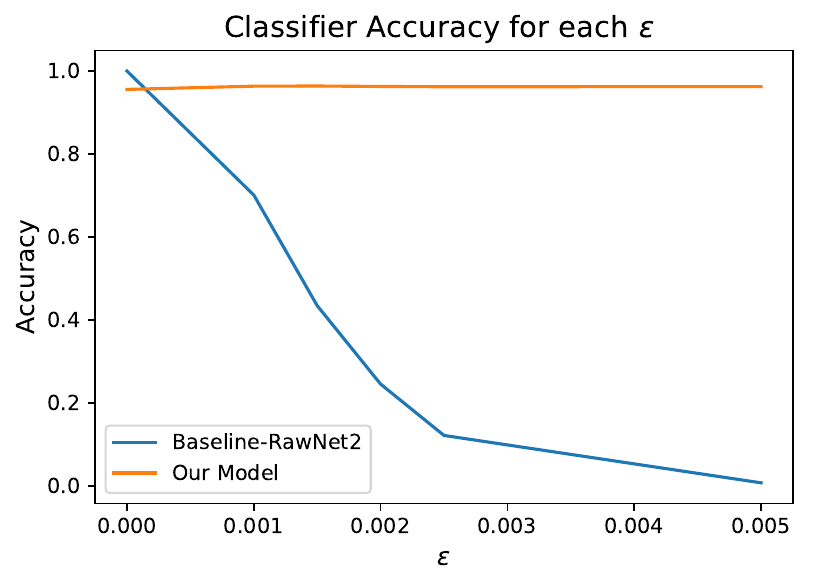}
			\caption{ 
				The accuracy for the RawNet2 (the best performing baseline) on adversarial samples generated using
				an $L_{\infty}$ attack. The
				adversarial samples decrease the accuracy of RawNet2 close to zero with minimum perturbations, while 
				our accuracy is minimally affect by the same samples. 
			} 
			\label{fig:adv_accuracy}
		\end{center}
	\end{figure}

We calculate the accuracy of RawNet2 against the samples at each value of $\epsilon$. 
For the smallest
$\epsilon$ of 0.001, the algorithm took an average of 22 iterations to find an adversarial sample. This continues
to decrease as we increase $\epsilon$ to 0.005, showing that our adversarial attack converges.
We see in Figure~\ref{fig:adv_accuracy} that RawNet2 decreases in accuracy to 0.7\% at $\epsilon=0.005$, a 99.3\% relative decrease in accuracy. These results call into question the robustness of previous ASVspoof results, and future competitions should directly incorporate adaptive adversaries as a means of more fully characterizing detector effectiveness. 
\\
\\ \noindent\fbox{\parbox{\columnwidth}{\textbf{Finding 3.} \emph{While other models which are susceptible to small changes in the sample (e.g., Gaussian noise), our models collects specific prosodic features from the audio which is not affected by such changes. This makes the approach of using prosody features for detection less susceptible to simple adaptive attacks. (\textbf{RQ2})}}}
\\
\begin{figure*}
	\begin{center}
		\includegraphics[width=\textwidth, keepaspectratio]{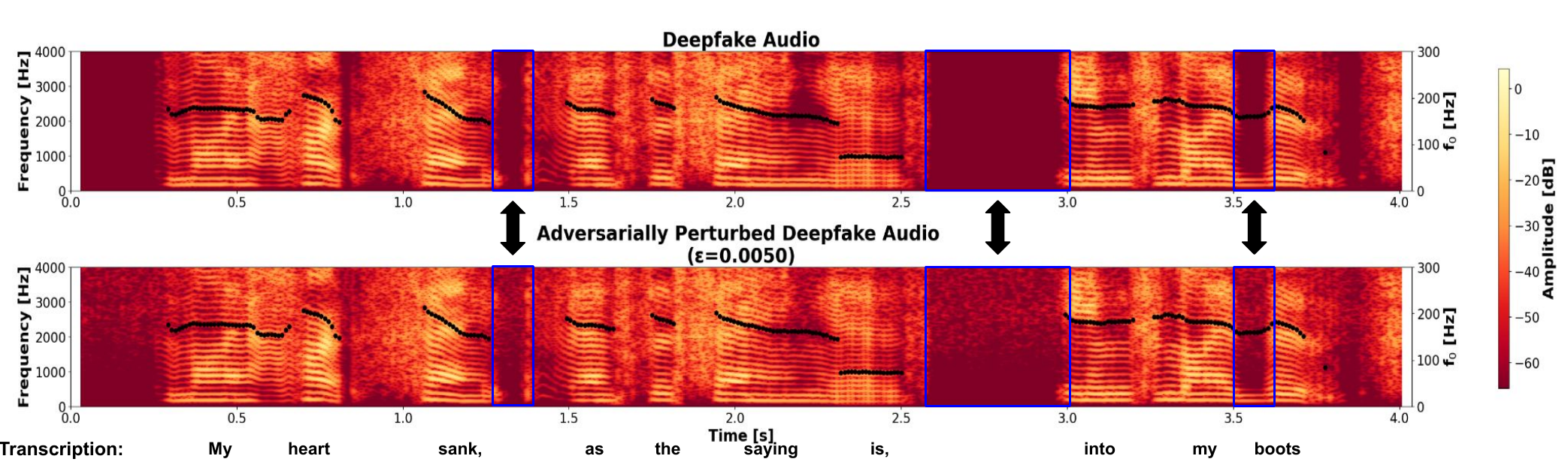}
		\caption{Examples of the spectrogram and fundamental frequency sequences for a deepfake 
		and the same deepfake modified by our adaptive adversary attack.
		Adversarial differences are highlighted above and illustrate the addition of white noise throughout the sample.
		However, the pitch information in black remains unchanged throughout the attack.} 
		\label{fig:two_spectrograms_adv}
	\end{center}
\end{figure*} 

\subsection{Lend Us Your Ear} \label{website} We strongly encourage the reader to visit our companion website.\footnote{https://sites.google.com/view/pitch-imperfect/home} The website contains two examples of 
deepfake audio and the targeted adversarial samples. The targeted audio samples have very little difference
between the original deepfake and the adversarial sample that RawNet2 misclassifies. This demonstrates how brittle
training on spectral features can be with just a minute amount of noise.
 A visual example of this phenomena
is demonstrated in Figure~\ref{fig:two_spectrograms_adv}.

\section{Discussion} \label{sec:disc} 

\subsection{Benefits of a Feature-Based Prosody Model}

As shown in Section~\ref{sec:adversarial}, there are several benefits over existing techniques when using a prosody feature detection approach. By preprocessing the input sample to extract the prosodic features, our model is resistent to optimization attacks on a sample. This requires the adversary to target the specific features that our model detects on (i.e., prosodic features), which as previously discussed in an ongoing field of research in both linguisitcs (e.g., understanding percieved prosody) and machine learning (e.g., properly injecting prosody into deepfakes). Additionally, by targeting the model at desired feature sets instead of allowing the model to freely train on input samples, we remove the ambiguity on the model's decision making. This is particularly important in cases where decisions have to be justified (e.g., removing content from social media or a business rejecting an incoming customer call) and is a necessary consideration for deploying these systems in industry. 

\subsection{Defense Robustness}
Current systems identify unnatural artifacts in deepfake audio files
such as spectral noise and distortion. As synthetic audio advances, these
generation problems will disappear and the quality of the audio file will also
improve. Unlike these systems, our approach is not dependent on artifacts in
the signal, but rather looks exclusively at the linguistic features of
speech itself. Research within the applied linguistics community demonstrates the difficulty of
determining which pitch variations are influential on our perceived
naturalness of speech~\cite{nooteboom1997prosody}. This makes training a
model to create natural-sounding pitch features not only a non-trivial task but
a significant open research challenge.

\subsection{Prosodic Edge Cases}
While we look at the high-level trends of prosody in human speech, there are
human voices that do not fall within the norm for prosodic features. Some
people naturally speak in a monotone voice or exhibit speech pathologies, making
their prosodic features differ from other humans and sometimes come across as
fake. Since our model looks solely at the prosody of a speaker, these unique
individuals could be misclassified as synthetic. 

\subsection{Data Limitations} \label{Dataset Limitations}

One of the largest limitations to the performance of our system is the type of data that is being produced for current datasets. Short two to five-second audio clips do not encapsulate how most people would experience a deepfake attack and limits the amount of prosodic information that is available. 
Weaponized deepfakes would resemble more of a conversation or a descriptive command rather than bursts of quick declarations. When more practical, lengthy deepfakes are compiled into datasets, our system will be able to pull more prosodic data from samples and better designate the linguistic trends in each sample. However, we use the ASVspoof2021 data for our experiments since that is the standard
currently used by the community.  

\section{Conclusion} \label{sec:conc} 
Advancements in audio deepfakes are making them increasingly indifferentiable and a growing concerns for not only the security community but also to the broader society. Current detection approaches are likely to become obsolete as deepfakes continue to become more realistic, which reinforces the need for the community to consider many different approaches to deepfake detection. In this paper, we evaluate the use of prosodic acoustic analysis as a means of detecting deepfakes and demonstrate that this approach contains comparable performance (e.g., 93\% accuracy and 24.7\% EER) to the currently baselines for deepfake detection. Additionally, we discuss the benefits to our prosody approach over the current baselines with the application of attention mechanisms for explainability and implementing the first adaptive adversary using an $L_\infty$ norm test to show the baseline's susceptibility to such attacks compared to our model. These results signify the beginning of exploration into alternative options for deepfake detection and a long-term effort to consider the intersection of linguistics and speech to protect against deepfake threats.

\bibliographystyle{IEEEtran}
\bibliography{disfluency}

\clearpage
\newpage

\appendix
\section{Acoustic Analysis Equations} \label{sec:meth:equations}

The following are the equations for jitter,
 shimmer, and Harmonic to Noise Ratio~\cite{teixeira_vocal_2013, praat_manual}, where $T_{i}$ is the period length, $A_{i}$ is the amplitude, and $N$ is the number of intervals.

\paragraph{Jitter local absolute}
The average absolute difference between consecutive 
periods, in seconds.
\begin{equation} \label{eq:jitt_abs}
    jitt_{abs}=\frac{1}{N-1}\sum_{i=1}^{N-1}\left | T_{i} - T_{i+1} \right |  
\end{equation}

\paragraph{Jitter local}
The average absolute difference between 
consecutive periods, divided by the average period.
\begin{equation} \label{eq:jitt_local}
    jitt = \frac{jitt_{abs}}{\frac{1}{N}\sum\limits_{i=1}^{N}T_{i}}\times 100
\end{equation}

\paragraph{Shimmer local}
The average absolute difference between the amplitudes 
of consecutive periods, divided by the average amplitude.
\begin{equation} \label{eq:shim_local}
    shim=\frac{\frac{1}{N-1}\sum\limits_{i=1}^{N-1}\left | A_{i}-A_{i+1} \right |}{\frac{1}{N}\sum\limits_{i=1}^{N}A_{i}}\times 100
\end{equation}

\paragraph{Harmonic to Noise Ratio}
Harmonic to Noise Ratio represents the degree 
of acoustic periodicity expressed in dB. 
$sig_{per}$ is the proportion of the signal that
is periodic. $sig_{noise}$ is the proportion of
the signal that is noise.
\begin{equation} \label{eq:hnr}
    HNR = 10 \times log_{10}\frac{sig_{per}}{sig_{noise}}
\end{equation}
\hfill

\section{Model Selection Architectures} \label{sec:model_selection_arch}

\begin{minipage}{\columnwidth}
    \fontsize{11}{9}\selectfont
    \begin{tabular}{|c|clccl|}
    \hline
                                     & \multicolumn{2}{c|}{Layer}   & \multicolumn{1}{c|}{Shape} & \multicolumn{2}{c|}{Parameters}                                                                   \\ \hline
    \multirow{8}{*}{Model A}         & \multicolumn{2}{c|}{LSTM}    & \multicolumn{1}{c|}{64}    & \multicolumn{2}{c|}{\begin{tabular}[c]{@{}c@{}}dropout=0.2\\ return\_sequences=True\end{tabular}} \\ \cline{2-6} 
                                     & \multicolumn{2}{c|}{BN}      & \multicolumn{1}{c|}{}      & \multicolumn{2}{c|}{}                                                                             \\ \cline{2-6} 
                                     & \multicolumn{2}{c|}{LSTM}    & \multicolumn{1}{c|}{32}    & \multicolumn{2}{c|}{dropout=0.2}                                                                  \\ \cline{2-6} 
                                     & \multicolumn{2}{c|}{BN}      & \multicolumn{1}{c|}{}      & \multicolumn{2}{c|}{}                                                                             \\ \cline{2-6} 
                                     & \multicolumn{2}{c|}{Dense}   & \multicolumn{1}{c|}{32}    & \multicolumn{2}{c|}{activation=ReLU}                                                              \\ \cline{2-6} 
                                     & \multicolumn{2}{c|}{Dropout} & \multicolumn{1}{c|}{}      & \multicolumn{2}{c|}{dropout=0.2}                                                                  \\ \cline{2-6} 
                                     & \multicolumn{2}{c|}{Dense}   & \multicolumn{1}{c|}{1}     & \multicolumn{2}{c|}{activation=sigmoid}                                                           \\ \cline{2-6} 
                                     & \multicolumn{5}{c|}{\begin{tabular}[c]{@{}c@{}}loss=binary\_crossentropy\end{tabular}}                                      \\ \hline
    \multirow{8}{*}{Model B}         & \multicolumn{2}{c|}{LSTM}    & \multicolumn{1}{c|}{100}   & \multicolumn{2}{c|}{\begin{tabular}[c]{@{}c@{}}dropout=0.2\\ return\_sequences=True\end{tabular}} \\ \cline{2-6} 
                                     & \multicolumn{2}{c|}{BN}      & \multicolumn{1}{c|}{}      & \multicolumn{2}{c|}{}                                                                             \\ \cline{2-6} 
                                     & \multicolumn{2}{c|}{LSTM}    & \multicolumn{1}{c|}{50}    & \multicolumn{2}{c|}{dropout=0.2}                                                                  \\ \cline{2-6} 
                                     & \multicolumn{2}{c|}{BN}      & \multicolumn{1}{c|}{}      & \multicolumn{2}{c|}{}                                                                             \\ \cline{2-6} 
                                     & \multicolumn{2}{c|}{Dense}   & \multicolumn{1}{c|}{50}    & \multicolumn{2}{c|}{activation=ReLU}                                                              \\ \cline{2-6} 
                                     & \multicolumn{2}{c|}{Dropout} & \multicolumn{1}{c|}{}      & \multicolumn{2}{c|}{dropout=0.2}                                                                  \\ \cline{2-6} 
                                     & \multicolumn{2}{c|}{Dense}   & \multicolumn{1}{c|}{1}     & \multicolumn{2}{c|}{activation=sigmoid}                                                           \\ \cline{2-6} 
                                     & \multicolumn{5}{c|}{\begin{tabular}[c]{@{}c@{}}loss=binary\_crossentropy\end{tabular}}                                      \\ \hline
    \multirow{12}{*}{Model C}        & \multicolumn{2}{c|}{LSTM}    & \multicolumn{1}{c|}{64}    & \multicolumn{2}{c|}{\begin{tabular}[c]{@{}c@{}}dropout=0.2\\ return\_sequences=True\end{tabular}} \\ \cline{2-6} 
                                     & \multicolumn{2}{c|}{BN}      & \multicolumn{1}{c|}{}      & \multicolumn{2}{c|}{}                                                                             \\ \cline{2-6} 
                                     & \multicolumn{2}{c|}{LSTM}    & \multicolumn{1}{c|}{32}    & \multicolumn{2}{c|}{dropout=0.2}                                                                  \\ \cline{2-6} 
                                     & \multicolumn{2}{c|}{BN}      & \multicolumn{1}{c|}{}      & \multicolumn{2}{c|}{}                                                                             \\ \cline{2-6} 
                                     & \multicolumn{2}{c|}{LSTM}    & \multicolumn{1}{c|}{16}    & \multicolumn{2}{c|}{dropout=0.2}                                                                  \\ \cline{2-6} 
                                     & \multicolumn{2}{c|}{BN}      & \multicolumn{1}{c|}{}      & \multicolumn{2}{c|}{}                                                                             \\ \cline{2-6} 
                                     & \multicolumn{2}{c|}{LSTM}    & \multicolumn{1}{c|}{8}     & \multicolumn{2}{c|}{dropout=0.2}                                                                  \\ \cline{2-6} 
                                     & \multicolumn{2}{c|}{BN}      & \multicolumn{1}{c|}{}      & \multicolumn{2}{c|}{}                                                                             \\ \cline{2-6} 
                                     & \multicolumn{2}{c|}{Dense}   & \multicolumn{1}{c|}{8}     & \multicolumn{2}{c|}{activation=ReLU}                                                              \\ \cline{2-6} 
                                     & \multicolumn{2}{c|}{Dropout} & \multicolumn{1}{c|}{}      & \multicolumn{2}{c|}{dropout=0.2}                                                                  \\ \cline{2-6} 
                                     & \multicolumn{2}{c|}{Dense}   & \multicolumn{1}{c|}{1}     & \multicolumn{2}{c|}{activation=sigmoid}                                                           \\ \cline{2-6} 
                                     & \multicolumn{5}{c|}{\begin{tabular}[c]{@{}c@{}}loss=binary\_crossentropy\end{tabular}}                                      \\ \hline
    \multirow{6}{*}{Model D}        & \multicolumn{2}{c|}{LSTM}    & \multicolumn{1}{c|}{100}   & \multicolumn{2}{c|}{\begin{tabular}[c]{@{}c@{}}dropout=0.2\\ return\_sequences=True\end{tabular}} \\ \cline{2-6} 
                                    & \multicolumn{2}{c|}{BN}      & \multicolumn{1}{c|}{}      & \multicolumn{2}{c|}{}                                                                             \\ \cline{2-6} 
                                    & \multicolumn{2}{c|}{Dense}   & \multicolumn{1}{c|}{100}   & \multicolumn{2}{c|}{activation=ReLU}                                                              \\ \cline{2-6} 
                                    & \multicolumn{2}{c|}{Dropout} & \multicolumn{1}{c|}{}      & \multicolumn{2}{c|}{dropout=0.2}                                                                  \\ \cline{2-6} 
                                    & \multicolumn{2}{c|}{Dense}   & \multicolumn{1}{c|}{1}     & \multicolumn{2}{c|}{activation=sigmoid}                                                           \\ \cline{2-6} 
                                    & \multicolumn{5}{c|}{\begin{tabular}[c]{@{}c@{}}loss=binary\_crossentropy\end{tabular}}                                      \\ \hline
    \multirow{8}{*}{Model E}        & \multicolumn{2}{c|}{LSTM}    & \multicolumn{1}{c|}{16}    & \multicolumn{2}{c|}{\begin{tabular}[c]{@{}c@{}}dropout=0.2\\ return\_sequences=True\end{tabular}} \\ \cline{2-6} 
                                    & \multicolumn{2}{c|}{BN}      & \multicolumn{1}{c|}{}      & \multicolumn{2}{c|}{}                                                                             \\ \cline{2-6} 
                                    & \multicolumn{2}{c|}{LSTM}    & \multicolumn{1}{c|}{16}    & \multicolumn{2}{c|}{dropout=0.2}                                                                  \\ \cline{2-6} 
                                    & \multicolumn{2}{c|}{BN}      & \multicolumn{1}{c|}{}      & \multicolumn{2}{c|}{}                                                                             \\ \cline{2-6} 
                                    & \multicolumn{2}{c|}{Dense}   & \multicolumn{1}{c|}{16}    & \multicolumn{2}{c|}{activation=ReLU}                                                              \\ \cline{2-6} 
                                    & \multicolumn{2}{c|}{Dropout} & \multicolumn{1}{c|}{}      & \multicolumn{2}{c|}{dropout=0.2}                                                                  \\ \cline{2-6} 
                                    & \multicolumn{2}{c|}{Dense}   & \multicolumn{1}{c|}{1}     & \multicolumn{2}{c|}{activation=sigmoid}                                                           \\ \cline{2-6} 
                                    & \multicolumn{5}{c|}{\begin{tabular}[c]{@{}c@{}}loss=binary\_crossentropy\end{tabular}}                                       \\ \hline
    \end{tabular}
\end{minipage}

\end{document}